\documentstyle[amssymb,12pt]{article}

\begin{document}

\title{NON-AUTONOMOUS SVINOLUPOV JORDAN KdV SYSTEMS}
\author{ Metin G{\" u}rses\\
{\small Department of Mathematics, Faculty of Sciences}\\
{\small Bilkent University, 06533 Ankara - Turkey}\\
%{\small and}\\
\\
\\
Atalay Karasu and  Refik Turhan  \\
{\small Department of Physics, Faculty of Arts and  Sciences}\\
{\small Middle East Technical University, 06531 Ankara-Turkey}}
\begin{titlepage}
\maketitle
\begin{abstract}
Non-autonomous Svinolupov -Jordan KdV systems are considered.
The integrability criteria of such systems are associated with
the existence of  recursion operators. A new non-autonomous KdV
system is obtained and its recursion operator is given for all $N$.
The examples for $N=2$ and $N=3$ are studied in detail. Some possible
transformations are also discussed which map some systems to 
autonomous cases.
\end{abstract}
\end{titlepage}
%\section{Introduction}
%{\Large
There is recently an increasing interest in the study of integrable nonlinear
partial differential equations on associative and non-associative
algebras \cite{sok1}  and in their recursion operators \cite{gur1},
 \cite{ping}. It is well known that one class of integrable autonomous
 multi-component  KdV equations (Korteweg-de Vries) , 
associated with  a Jordan algebra $J$ (commutative and non-associative),

\begin{equation}
 q^{i}_{t}=q^{i}_{xxx}+ s^{i}_{j k}q^{j}q^{k}_{x} \,\,\,\,\,
, s^{i}_{jk}=  s^{i}_{kj},  \,\,\,\ i,j,k=1,2,...,N
 \label{a0}
\end{equation}

\noindent
has been considered by Svinolupov \cite{SVI} where
$q^{i}$ are real and depend on the variables $x$
and ${t}$. The constant parameters
 $ s^{i}_{jk}$  are structure constants ,with respect to some basis
$e_{i}$ ,of a Jordan algebra $J$

\begin{equation}
  e_{i}\circ e_{j}= s^{k}_{ij}e_{k}
\end{equation}
satisfy the commutativity identities

\begin{equation}
 s^{k}_{p\,r} F^{i\,\,\,}_{ljk}+ s^{k}_{j\,r}
 F^{i\,\,\,}_{lpk}
+ s^{k}_{j\,p} F^{i\,\,\,}_{lrk}=0, \label{a1}
\end{equation}

\noindent
where

\begin{equation}
 F^{i\,\,\,}_{plj}=  s^{i}_{j\,k} s^{k}_{l\,p}
- s^{i}_{l\,k} s^{k}_{j\,p} ,  \label{a2}
\end{equation}
is the associator of the Jordan algebra.
The integrability criteria of the multi-component Jordan KdV system
(JKdV)(\ref{a0}) are associated with the existence of higher symmetries and
the corresponding recursion operator.

\vspace{0.3cm}

\noindent
{\bf Theorem 1:} (Svinolupov)\, {\it Let $ s^{i}_{jk}$ be the structure constants of a Jordan
algebra,i.e.,satisfy the identities (\ref{a1}). The system (\ref{a0})
possesses a recursion operator of the form}

\begin{equation}
 {\cal R}^{i}_{j}=\delta^{i}_{j} D^{2}+{2 \over3}\, s^{i}_{jk}\,q^{k}
+{1 \over 3}\,s^{i}_{jk}\,q^{k}_{x}\,D^{-1}
+{1 \over 9}\,\,( s^{i}_{jm} s^{m}_{kl}-
 s^{i}_{km} s^{m}_{jl})\,q^{l}\,D^{-1}\,q^{k}\,D^{-1}.
\end{equation}

\vspace{0.3cm}

\noindent
We only need to prove that  ${\cal R}$ satisfies the integrability
condition \cite{olv}

\begin{equation}
{\cal R}^{i}_{j , t}= \,K^{\prime \, i}_{k}\,\,{\cal R}
^{k}_{j}\,\,-\,\,{\cal R}^{i}_{k}\,\,K^{\prime \,k}_{j} ,    \label{a5}
\end{equation}

\noindent
with respect to (\ref{a1})
where $K^{\prime \, i}_{k}$ is the Fr{\'e}chet derivative of the
system (\ref{a0}). Therefore,the existence of the recursion operator
ensures that the system (\ref{a0}) possesses an infinite series
of symmetries.

Svinolupov established a one to one  correspondence between Jordan
algebras and the subsystems(reducible,irreducible,completely reducible)
of the system (\ref{a0}).

\vspace{0.3cm}

\noindent
{\bf Definition 1:}\, A system of type (\ref{a0}) is called reducible
(triangular) if it decouples into the form

\begin{eqnarray}
&&U^{i}_{t}=F^{i}(U^{k},U^{k}_{x},U^{k}_{xxx}),~~~~~~~~
i,k=1,2,...,K, ~~~~~~~~0<K<N\\
&&V^{i}_{t}=G^{i}(U^{k},U^{k}_{x},V^{k},V^{k}_{x},V^{k}_{xxx}),~~~~~~~~
i=1,2,...,N-K.
\end{eqnarray}

\noindent
under a certain linear transformation which leaves the system
(\ref{a0}) invariant. If not,it is irreducible. A system is called
completely reducible if the second equation given above does not contain the
dynamical variables $U^{i}$  and $U^{i}_{x}$.

\vspace{0.3cm}

\noindent
{\bf Example 1:}\, For $N=2$,the complete classification,with respect to
Jordan algebra, was given by Svinolupov \cite{SV1}.

\begin{eqnarray}
&& u_{t}=u_{xxx}+ 2c_{0}uu_{x},~~~~~~
 v_{t}= v_{xxx}+ c_{0}(uv)_{x}, \label{c0} \\
&& u_{t}=u_{xxx}+ c_{0}uu_{x},~~~~~~~
 v_{t}= v_{xxx}+ c_{0}(uv)_{x}, \label{c1}  \\
&& u_{t}=u_{xxx},~~~~~~~~~~~~~~~~~~~
 v_{t}= v_{xxx}+ c_{0}\,uu_{x}, \label{c2}
\end{eqnarray}

\noindent
where $c_{0}$ is an arbitrary constant.The reducible systems (\ref{c0})
and (\ref{c1}) correspond to the JKdV and trivially JKdV (associator is zero)
respectively.The last system is completely reducible system. 

\vspace{0.3cm}

\noindent
{\bf Example 2:}For $N=3$,

\noindent
i)The system

\begin{eqnarray}
&& u_{t}=u_{xxx}- c_{0}(u^{2}-v^{2}-w^{2})_{x},\nonumber\\
&& v_{t}= v_{xxx}-c_{0}(uv)_{x},\nonumber\\
&& w_{t}= w_{xxx}-c_{0}(uw)_{x} \label{c3}
\end{eqnarray}

\noindent
is the only irreducible JKdV system \cite{SV1},\cite{ator}.

\noindent
ii) A reducible JKdV system is

\begin{eqnarray}
&& u_{t}=u_{xxx}- 2c_{0} uu_{x},\nonumber\\
&& v_{t}= v_{xxx}-c_{0}(uv)_{x},\nonumber\\
            && w_{t}= w_{xxx}- c_{0}(uw)_{x}. \label{c4}
\end{eqnarray}

\noindent
In this work we will investigate the non-autonomous Svinolupov
JKdV systems. For this purpose,  we consider the non-autonomous form of the
system (\ref{a0}) as

\begin{equation}
 q^{i}_{t}=q^{i}_{xxx}+s^{i}_{j k}(t)q^{j}q^{k}_{x} ,~~~~~
s^{i}_{jk}(t)= s^{i}_{kj}(t),  \,\,\,\ i,j,k=1,2,...,N
\label{a3}
\end{equation}

\noindent
where $s^{i}_{j k}(t)$ are sufficiently differentiable functions.
In particular,for $N=1$ the system (\ref{a3}) is the well known
 cylindrical KdV (cKdV) equation \cite{Cal}

\begin{equation}
u_{t}=u_{xxx}+{6 \over \sqrt {t}}uu_{x},
\end{equation}

\noindent
which possesses a recursion operator \cite{OEV}

\begin{equation}
{\cal R}= tD^{2}+4 \sqrt{t}u +{1 \over 3}x
+ {1 \over 6}(12 \sqrt{t}u_{x}+1)D^{-1}.
\end{equation}

\noindent
We are now in a position to propose a recursion operator
for the integrability of the system (\ref{a3}). Moreover,
motivated  by the results obtained in Refs.(\cite{SVI}, \cite{SV1}) and
(\cite{OEV}-\cite{kar1}) we may state the following theorem.

\vspace{0.3cm}

\noindent
{\bf Theorem 2:}\, {\it  Let} $s^{i}_{jk}$ {\it be the structure constants 
of a Jordan algebra,i.e.,satisfy the identities (\ref{a1}). The system 
(\ref{a3}) possesses a recursion operator of the form}.

\begin{eqnarray}
{\cal R}^{i}_{j}&=& t \delta^{i}_{j}\,D^{2} +{2 \over 3}\,\sqrt{t}
\, s^{i}_{jk}\,q^{k}+ {1 \over 3}\delta^{i}_{j}x
+({1 \over 3}\,\sqrt{t}\, s^{i}_{jk}\,q^{k}_{x}+{1 \over 6}
  \delta^{i}_{j}) \,D^{-1} \nonumber \\
&&  +{1 \over 9}\,F^{i}_{\,\,lkj}\,q^{l}\,D^{-1}\,q^{k}\,D^{-1}. \label{a6}
\end{eqnarray}

\noindent
We only need to prove that  ${\cal R}$ satisfies the integrability
condition (\ref{a5}) with respect to (\ref{a1}).

\vspace{0.3cm}

\noindent
{\bf Example 3:}\, For $N=2$,

\noindent
i)The system

\begin{eqnarray}
&& u_{t}=u_{xxx}+ {2c_{0} \over \sqrt {t}}uu_{x},\nonumber\\
&& v_{t}= v_{xxx}+ {c_{0} \over \sqrt {t}}(uv)_{x}, \label{a7}
\end{eqnarray}

\noindent
is the non-autonomous JKdV where $c_{0}$ is an arbitrary constant.
The  recursion operator ${\cal R}$ for the above system is

\begin{equation}
{\cal R}=
\left(
\begin{array}{cc}
{\cal R}^{0}_{0}
 & {\cal R}^{0}_{1} \\
{\cal R}^{1}_{0}  &
{\cal R}^{1}_{1}
\end{array} \; \; \right)\;.
  \label{a03}
\end{equation}

\noindent
with

\begin{eqnarray}
{\cal R}^{0}_{0} & = & tD^{2}+{1 \over 3}x+{4c_{0} \over 3}\sqrt{t}u
+ {1 \over 6}(4c_{0}\sqrt{t}u_{x}+1)D^{-1},
 \nonumber \\
{\cal R}^{0}_{1} & = & 0,
\nonumber\\
{\cal R}^{1}_{0} & = &
 {2c_{0} \over 3}\sqrt{t}v + {c_{0} \over 3}\sqrt{t}v_{x}D^{-1}
-{c_{0}^{2} \over 9} uD^{-1}vD^{-1}, \nonumber \\
{\cal R}^{1}_{1} & = &
tD^{2}+{1 \over 3}x+{2c_{0} \over 3}\sqrt{t}u+ {1 \over 6}(2 c_{0}\,
\sqrt{t}u_{x}+1)D^{-1}+{c_{0}^{2} \over 9}uD^{-1}uD^{-1}.
\label{z1}
\end{eqnarray}

\noindent
ii) The non-autonomous reducible JKdV is

\begin{eqnarray}
&& u_{t}=u_{xxx}+ {c_{1} \over \sqrt {t}}uu_{x},\nonumber\\
&& v_{t}= v_{xxx}+ {c_{1} \over \sqrt {t}}(uv)_{x}, \label{a8}
\end{eqnarray}

\noindent
where $c_{1}$ is an arbitrary constant.
The recursion operator for this system is

\begin{eqnarray}
{\cal R}^{0}_{0} & = & tD^{2}+{1 \over 3}x+{2c_{1} \over 3}\sqrt{t}u
+ {1 \over 6}(2c_{1}\sqrt{t}u_{x}+1)D^{-1},
 \nonumber \\
{\cal R}^{0}_{1} & = & 0,
\nonumber\\
{\cal R}^{1}_{0} & = &
 {2c_{1} \over 3}\sqrt{t}v + {c_{1} \over 3}\sqrt{t}v_{x}D^{-1},
 \nonumber \\
{\cal R}^{1}_{1} & = &
tD^{2}+{1 \over 3}x+{2c_{1} \over 3}\sqrt{t}u+ {1 \over 6}(2c_{1}\sqrt{t}u_{x}
+1)D^{-1}.
\label{z2}
\end{eqnarray}

\vspace{0.3cm}

\noindent
{\bf Example 4:}\, For $N=3$

\noindent
i) The non-autonomous irreducible JKdV system is

\begin{eqnarray}
&& u_{t}=u_{xxx}- {c_{0} \over \sqrt {t}}(u^{2}-v^{2}-w^{2})_{x},\nonumber\\
&& v_{t}= v_{xxx}-{c_{0} \over \sqrt {t}}(uv)_{x},\nonumber\\
&& w_{t}= w_{xxx}-{c_{0} \over \sqrt {t}}(uw)_{x}. \label{a9}
\end{eqnarray}

\noindent
ii)The non-autonomous reducible J KdV system

\begin{eqnarray}
&& u_{t}=u_{xxx}- {2c_{0} \over \sqrt {t}}uu_{x},\nonumber\\
&& v_{t}= v_{xxx}-{c_{0} \over \sqrt {t}}(uv)_{x},\nonumber\\
&& w_{t}= w_{xxx}-{c_{0} \over \sqrt {t}}(uw)_{x}, \label{a10}
\end{eqnarray}

\noindent
is the extension of (\ref{c0}).
The recursion operators for the systems  (\ref{a9}) and (\ref{a10})
are too long,hence we do not give them here. 

Finally, we establish linear transformations between
autonomous and non-autonomous systems. In the scalar case,
the KdV and cKdV equations are equivalent since their solutions
are related by simple Lie-point transformation \cite{blum}-\cite{FUC}.

\begin{equation}
u(x,t)=t^{-1/2}u^{\prime}(xt^{-1/2},-2t^{-1/2})
-{1 \over 12}xt^{-1/2}.
\end{equation}

\noindent
Here we present a generalization of this result to the case
of systems of evolution equations.

\vspace{0.3cm}

\noindent
{\bf Definition 2:} Two systems of equations

\begin{eqnarray}
&&u^{i}_{t}= u^{i}_{xxx}+ f(x,t,u^{i},u^{i}_{x}),\nonumber \\
&&u^{\prime i}_{\sigma}= u^{\prime i}_{\xi\xi\xi}
+ g(\xi,\sigma,u^{\prime i},u^{\prime i}_{\xi}),
\end{eqnarray}

\noindent
will be called equivalent if there exists a change of variables
of the form

\begin{eqnarray}
&&\xi=\alpha(t)x+\beta(t),~~~~~~\sigma=\gamma(t), \nonumber\\
&&u^{i}(x,t)=\Gamma(t)u^{\prime i}(\xi(x,t),\sigma(x,t))+\eta(x,t), 
\label{d2}
\end{eqnarray}

\noindent
which is invertible. The first result is given in the following statement.

\vspace{0.3cm}

\noindent
{\bf Proposition 1:}\, {\it The system

\begin{eqnarray}
&& u_{t}=u_{xxx}+ {c_{0} \over \sqrt {t}}uu_{x},\nonumber\\
&& v_{t}= v_{xxx}+ {c_{1} \over \sqrt {t}}(uv)_{x},
\end{eqnarray}

\noindent
where $c_{0}$ and $c_{1}$ arbitrary constants, may be transformed
into the autonomous perturbation of KdV system

\begin{eqnarray}
&&u^{\prime}_{\sigma}=u^{\prime}_{\xi \xi \xi}
+ c_{0}u^{\prime}u^{\prime}_{\xi},\nonumber\\
&& v^{\prime}_{\sigma}= v^{\prime}_{\xi\xi\xi}
+c_{1}(u^{\prime}v^{\prime})_{\xi}, \label{d0}
\end{eqnarray}

\noindent
through a transformation of the form (\ref{d2})
if and only if $ c_{0} = c_{1}$.
}
\vspace{0.3cm}

\noindent
The validity of this Proposition allows us to state the following.

\vspace{0.3cm}

\noindent
{\bf Proposition 2:}{\it 
The non-autonomous JKdV system (\ref{a8}) is transformed into
the autonomous JKdV system(\ref{c1}) through the transformation of the form

\begin{eqnarray}
 &&u(x,t)=t^{-1/2}u^{\prime}(xt^{-1/2},-2t^{-1/2})
-{1 \over 2c_{1}}xt^{-1/2},\nonumber \\
 && v(x,t)=t^{-1/2}v^{\prime}(xt^{-1/2},-2t^{-1/2}) . \label{d3}
\end{eqnarray}
}
\vspace{0.3cm}

\noindent
Similar to Propositions 1 and 2 we have the following statement.

\vspace{0.3cm}

\noindent
{\bf Proposition 3:} {\it The non-autonomous JKdV system (\ref{a9})
 is transformed into the
the autonomous JKdV system (\ref{c3}) through the transformation

\begin{eqnarray}
 &&u(x,t)=t^{-1/2}u^{\prime}(xt^{-1/2},-2t^{-1/2})+
{1 \over 4c_{0}}xt^{-1/2},\nonumber\\
&&v(x,t)=t^{-1/2}v^{\prime}(xt^{-1/2},-2t^{-1/2}), \nonumber\\
&&w(x,t)=t^{-1/2}w^{\prime}(xt^{-1/2},-2t^{-1/2}).
\end{eqnarray}
}

\vspace{0.3cm}

From the above discussions  we have the following result.

\vspace{0.3cm}

\noindent
{\bf Proposition 4:} {\it The non-autonomous JKdV system  (\ref{a7})
(or its extension (\ref{a10})) can not be transformed into the JKdV 
system (\ref{c0}) (or its extension (\ref{c4})) through a transformation 
of the form (\ref{d2}).}

\vspace{0.3cm}

We have observed that for some special cases of $N=2$ and $N=3$ time
dependent systems transform to time independent cases. This comes indeed
from the type of the Jordan algebra. For general $N$ we have the following
statement

\vspace{0.3cm}

\noindent
{\bf Proposition 5.}\, {\it A Jordan system (\ref{a3}) is equivalent to 
an autonomous Jordan system (\ref{a0}) if there exists an element {\bf a}
of $J$
such that $ {\bf a}^2={\bf a}$ and $q \circ {\bf a}=q$ for all $q \in J$}

\vspace{0.3cm}

\noindent
{Proof.}\, We write the system of equations (\ref{a3}) in the form
$q_{t}=q_{xxx}+{1 \over \sqrt{t}}\, q \circ q_{x}$., where $q$  takes
values in a Jordan algebra $J$. Take the point transformation 

\begin{eqnarray}
q(x,t)&=&t^{-1/2}\, v(\xi, \tau)-{1 \over 2}\,xt^{-1/2}\, {\bf a} \nonumber\\
\xi&=&xt^{-1/2},~~~~~\tau=-2t^{-1/2}.
\end{eqnarray}

\noindent
Then equations for $v$ becomes time independent.

\vspace{0.3cm}

\noindent
Transformable case in $N=2$ (Example (2.ii)) is the case  
with ${\bf a}=e_{1}$ where $\{e_{i}, 
i=1,2 \}$ are a basis of $J$. The Example (4.i) in $N=3$ case 
is also transformable because the element ${\bf a}= -{1 \over 2c_{0}}\, e_{1}$
satisfies the condition ${\bf a}^2={\bf a}$.

We would like to remark on the symmetries of (\ref{a3}). The first symmetry
is the $x$-translational symmetry $\sigma^{i}_{1}=q^{i}_{x}$. 
The next one is the
scale symmetry $\sigma^{i}_{2}=t\,q^{i}_{t}+{1 \over 3}\,x q^{i}_{x}+
{1 \over 6}\,q^{i}$. The first generalized symmetry is given by
$\sigma^{i}_{3}= {\cal R}^{i}_{j}\, \sigma^{j}_{2}$, where ${\cal R}$ is the
recursion operator (\ref{a6}) of the system (\ref{a3}). This symmetry is
nonlocal and contains the associator (tensor $F^{i}_{jkl}$) of the algebra $J$.
There exists also an additional symmetry, the Galilean symmetry,
 $\eta^{i}_{1}= \sqrt{t}\, S^{i}_{jk}\,q^{k}_{x}\,k^{j}+{1 \over 2} k^{i}$
 for the system (\ref{a3}) satisfying $S^{i}_{jk}\, k^{j}=\delta^{i}_{k}$.  
Here we remark also that the element ${\bf k}=k^{i}\, e_{i}$ of $J$
 satisfies ${\bf k}^2={\bf k}$ hence due to the above proposition 5
the corresponding systems are transformable to autonomous KdV systems 
(\ref{a0}). In the general case, $F \ne 0$, 
$\sigma^{i}=\Lambda^{i}_{j}\,k^{j}$ is a symmetry of the non-autonomous 
JKdV system (\ref{a3}) for all $k$, where 

\begin{equation}
\Lambda^{i}_{j}={1 \over 3}\, \sqrt{t}\, s^{i}_{jk}\,q^{k}_{x}+
{1 \over 6}\, \delta^{i}_{j}+{1 \over 9}\,F^{i}_{lkj}\,q^{l} D^{-1}\,q^{k}.
\end{equation}

In the case of time dependent recursion operators (and time dependent
evolution equations) there is an ambiguity in calculating the symmetries.
It is claimed that the recursion operators do not in general map symmetries
to symmetries \cite{sw}. Following \cite{sw}
time dependent higher symmetries can be constructed recursively by means of
the extended recursion operator 

\begin{equation}
{\cal R}_{e}={\cal R}+\Lambda \, \int^{t}\, dt^{\prime}\, \Pi D^2,
\end{equation}

\noindent
where ${\cal R}$ is the recursion operator given in (\ref{a6}), 
and $\Pi$ is the projection on the kernel of $D$ defined explicitly by
$\Pi f(t,x,q,q_{x}, \cdots)=f(t,0,0,0, \cdots)$.

\vspace{0.3cm}

This work is partially supported by the Scientific and Technical
Research Council of Turkey (TUBITAK) and  Turkish Academy of Sciences (TUBA).

\end{document}